# Network-based brain computer interfaces: principles and applications

**Juliana Gonzalez-Astudillo[1,2], Tiziana Cattai[1,2,3], Giulia Bassignana[1,2], Marie-Constance Corsi[1,2] and Fabrizio De Vico Fallani[1,2,*]**

[1] Inria Paris, Aramis Project Team, Paris, France

[2] Institut du Cerveau, ICM, Inserm U 1127, CNRS UMR 7225, Sorbonne Université, Paris, France

[3] Dept. of Information Engineering, Electronics and Telecommunication, University Sapienza, Rome, Italy

*Corresponding author: fabrizio.de-vico-fallani@inria.fr

## Abstract

Brain-computer interfaces (BCIs) make possible to interact with the external environment by decoding the mental intention of individuals. BCIs can therefore be used to address basic neuroscience questions but also to unlock a variety of applications from exoskeleton control to neurofeedback (NFB) rehabilitation. In general, BCI usability critically depends on the ability to comprehensively characterize brain functioning and correctly identify the user's mental state. To this end, much of the efforts have focused on improving the classification algorithms taking into account localized brain activities as input features. Despite considerable improvement BCI performance is still unstable and, as a matter of fact, current features represent oversimplified descriptors of brain functioning. In the last decade, growing evidence has shown that the brain works as a networked system composed of multiple specialized and spatially distributed areas that dynamically integrate information. While more complex, looking at how remote brain regions functionally interact represents a grounded alternative to better describe brain functioning. Thanks to recent advances in network science, i.e. a modern field that draws on graph theory, statistical mechanics, data mining and inferential modelling, scientists have now powerful means to characterize complex brain networks derived from neuroimaging data. Notably, summary features can be extracted from these networks to quantitatively measure specific organizational properties across a variety of topological scales. In this topical review, we aim to provide the state-of-the-art supporting the development of a network theoretic approach as a promising tool for understanding BCIs and improve usability.

## 1. Introduction

Over the past decades, the way scientists have looked at the human brain has witnessed a paradigm shift. The view that cognition and behavior result from localized neuronal ensembles has progressively left room for the realization that their interaction is what really matters. Today, we know that the brain is not just a collection of isolated units working independently, but it rather consists of a complex network that integrates information across differently specialized regions via anatomical as well as functional connections [1].

Such transition from a reductionist to a holistic perspective has been accompanied by the dawning of network science, i.e. a modern field drawing on graph theory that summarizes and quantifies organizational properties of complex interconnected systems. In human neuroscience, brain regions are treated as network nodes, and the connections between the nodes - inferred from structural or functional neuroimaging data - are represented as network edges (or links) [2–4]. Network properties including efficiency [5], modularity [6], and node centrality [7] have been demonstrated to support basic cognitive functions such as language and memory [1]. Critically, these network indexes are also sensitive to physiological and pathological alterations of the mental state and can capture brain organizational mechanisms across different spatiotemporal scales [8,9].

Such fundamental relationship between network topology and brain function is a key element of modern neuroscience and offers a grounded tool for analyzing brain networks by means of few topological descriptors rather than high-dimensional connectivity matrices [10]. Network neuroscience has allowed answers to fundamental questions spanning consciousness, plasticity, and learning, but it can also play a role in engineering applications aiming to characterize different brain states and recognize mental intentions from functional neuroimaging recordings. This is the case of brain-computer interfaces (BCIs) - also referred to as brain-machine interfaces - which implement ideal communication pathways bypassing the traditional effector of the musculoskeletal system and directly interacting with external devices [11–13]. Based on the classification of mental states from brain activity, BCIs are increasingly explored for control and communication [13–15], and for treatment of neurological disorders [16,17].

In this context, the first findings have shown that the modulation of brain activity elicited by motor imagery (MI) [18] as well as by decision-making tasks [19] generates detectable signal changes, such as power spectrum density (PSD), localized in one or few brain sites. These features are still surprisingly used to develop modern BCIs despite the limited performance and poor usability. However, examining the signal of one specific region while neglecting its interactions with other regions, oversimplifies the real phenomenon and one must instead obtain an understanding of the system's collective behavior to fully capture the brain functioning. This can be in part explained by the fact that the BCI community has mainly focused on improving the signal processing and classification block of the BCI pipeline, while neglecting the feature extraction part [20]. Hence, the research of new features represents a fertile field of research as witnessed by the increasing number of studies adopting network-based perspectives in an effort to understand and improve BCI performance [21].

Here, we provide a topical review that describes how to construct functional brain networks, surveys network theoretic measures, and illustrates their application to cognitive and motor BCI-related neuroimaging data (**Figure 1**). This review can be mainly thought of as a methodological reference and does not aim to provide new neurophysiological insights. Throughout the sections, we comment on the methodological limitations, the best practices for their application, and possible future directions. This review is written for the neural engineering community, and so the literature we cover and the examples that we present are selected to be especially relevant for researchers working with electroencephalography (EEG), magnetoencephalography (MEG) or functional magnetic resonance imaging (fMRI) data. Our goal is to provide an accessible introduction to the field, and to inspire the younger generation of scientists willing to study BCIs through network neuroscience.

## 2. From functional neuroimaging data to brain networks

The first step when studying brain networks is to decide which are the nodes and the edges. Typically, the definition of the nodes depends on the specific neuroimaging modality. For fMRI and other voxel-based techniques, the most common approach consists in using anatomical atlas and each region of interest (ROI) corresponds to a node [22,23]. For sensor-based modalities, such as EEG and MEG, each sensor typically corresponds to a node [4,24], although source-reconstruction techniques can be used to define nodes at the cortical level [25–27]. Because neuroimaging techniques only give access to regional activities, recorded as signals, the network edges must be inferred using statistical approaches. This is typically done by means of functional connectivity (FC) estimators which measure the temporal dependency between different brain signals. As a result, network edges correspond to FC estimates.

In the last decades, many methods have been developed to quantify functional interactions in the brain, relying on tools from signal processing and information theory. Even if each method is characterized by its own specific operations, the general procedure remains the same. Given a set of time series corresponding to the activity of different brain sites, the goal is to quantify the interaction between every signal pair. The literature is consistent in recognizing that the first distinction between FC estimators is between undirected and directed methods [3,28]. The former measures symmetric interactions, without considering the directionality of the information flow. The latter characterizes causal effects during activity propagation. Inside these categories, further distinctions can be done, according to their ability to describe linear or nonlinear interactions, bivariate or multivariate effect, time or frequency domain properties. **Table 1** shows a non-exhaustive list of the most used FC estimators in neuroscience, with their associated properties. In the following, we present some of the most challenging issues that significantly influence connectivity estimation.

### 2.1. Critical aspects

#### 2.1.1. Spurious connectivity

The ultimate goal of FC methods is to quantify true signal interactions between different brain areas. However, several conditions can affect the correct estimation and introduce spurious contributions, thus giving a potentially distorted measure of the real interactions. This is in part due to the fact that most of the experimental techniques for recording noninvasve human brain signals, such as EEG, MEG or fMRI [29–31] can only indirectly capture the real neuronal source activity. For example, EEG and MEG measure respectively the electrical activity and magnetic flux produced by neurons within the brain. The electromagnetic signals propagate through the head tissues from the cortex - i.e., the source space - to the scalp - i.e., the sensor space. During this propagation, the different electrical conductibility of the tissues generates a spatial smearing of the signals on the scalp [32,33]. As a consequence, the signal measured in one electrode does not reflect the activity of one single source and this phenomemon, also known as volume conduction effect, can lead to spurious instantaneous interactions [34]. One possible solution consists of computing FC in the source domain, after having reconstructed the signals of the cortical space by means of inverse procedures. While source reconstruction techniques do alleviate the volume conduction effect, they do not entirely solve the problem and results can strongly depend on the implemented algorithm [35]. Furthermore, individual head models obtained from structural MRI are often necessary to have best high-quality results [26,27].

| Functional connectivity estimators | | Properties | | |
|---|---|---|---|---|
| | | *Non-linearity* | *Time-varying* | *Multivariate* |
| *Undirected* | Spectral coherence [42] | - | - | - |
| | Imaginary coherence [34] | - | - | - |
| | Phase-Locking Value [43] | ✓ | - | - |
| | Weighted phase lag index [44] | ✓ | - | - |
| | Partial coherence [39] | - | - | ✓ |
| | Synchronization likelihood [45] | ✓ | - | - |
| | Mutual information [46] | ✓ | - | - |
| | Wavelet coherence [47] | ✓ | ✓ | |
| *Directed* | Granger causality [48] | - | - | - |
| | Kernel Granger causality [49] | ✓ | - | - |
| | Partial Granger causality [50] | - | - | ✓ |
| | Partial directed coherence [41] | - | - | ✓ |
| | Transfer Entropy [46] | ✓ | - | ✓ |
| | Directed Transfer Function [51] | ✓ | - | ✓ |
| | Adaptive partial directed coherence [52] | - | ✓ | ✓ |

**Table 1- Selection of the most commonly used functional connectivity (FC) estimators**. The different methods are organized according to their ability to capture directed or undirected interactions. Specific properties associated with some of the critical issues discussed in the section are reported on the right part of the table.

Because volume conduction effects exclusively affect instantaneous interactions, an alternative solution is the use of FC estimators that purposely remove lag-zero contributions from the estimates, such as *imaginary coherence* [34], or *weighted phase lag index*. While these approaches significantly limit the bias introduced by the volume conduction smearing, they might remove possibly existing instantaneous neurophysiological signal interactions [36]. Spurious FC can also be introduced by third-party influences when multiple signals are available. When estimating the interaction between two signals, a portion of the interaction might be merely given by the presence of a third signal interacting with them. In some cases, it is therefore crucial to isolate this contribution and eventually remove it from the estimate [37,38]. While the large part of the FC methods have tended to neglect third-party influences, there are now several methods in literature, based on *partial coherence* [39,40] or *partial directed coherence* [41], which have been designed to circumvent and alleviate those spurious effects.

### 2.1.2. Non-linear interactions

The neural system at a microscopic scale is characterized by nonlinear dynamics such as those of neuronal responses to stimuli or synaptic transmission [53]. A crucial question is whether the brain activity at a macroscopic scale can be instead approximated by linear dynamics and take advantage of the efficacy of linear methods [54]. The findings related to this subject are controversial [55]. Several studies have investigated nonlinearities in brain signals using the largest Lyapunov exponent, the correlation integral or the method of data surrogate. The obtained results show that in healthy subjects there is a weak signal nonlinearity [56,57]. Other works have reported nonlinear behavior in epileptic patients explained by the transitions between ordered and disordered states and the low-dimensional chaos [58,59]. The latter evidence was nevertheless contradicted by more recent endeavors showing that even in diseased subjects, nonlinear methods perform as well as linear ones [60,61].

More in general, nonlinearity also concerns the statistical interdependence between different brain signals. This typically means that FC is not proportional to either magnitude or phase of the signal frequency contents. In the early 1980s, the concept of synchronization was already extended and explained as a result of the adjustment of the oscillators caused by the presence of weak interactions [62]. In these situations, the use of linear FC can fail to provide a complete description of the temporal properties of the signal interactions.Despite such limitation, the majority of FC studies still rely on linear-based interaction methods because of their simplicity and intuitive interpretation. In the case of *spectral coherence* and related estimators (*partial coherence*, *imaginary coherence*, etc..) it has been shown that they are relatively robust to nonlinear fluctuations in the signal amplitudes but not in phases [63]. However, if there is a precise for nonlinearity, several estimators can be used to capture nonlinear FC taking into undirected (*mutual information* [64], *phase locking value*, *synchronization likelihood*) or directed relationships (e.g., *transfer entropy* [65], *kernel Granger-causality*) (**Table 1**).

### 2.1.3. Time-varying dynamics

FC estimators have been typically applied to extract connectivity patterns characterizing relatively long time periods (from dozens of seconds to minutes). In the last decade, the focus has shifted to shorter time scales that can be studied with dynamic functional connectivity (dFC) [66]. Indeed, the possibility to determine how FC fluctuates during specific tasks is particularly appealing for BCI applications, where the mental state of the subjects rapidly varies to control the effector or accommodate the feedback. To this end, the simplest approach consists of reducing the length of the time window, letting it slide along the entire period of interest, with or without overlapping. On the one hand, reducing the size of the time window has also the effect of ensuring the signal (quasi)stationarity hypothesis required by many FC estimators [67–69]. On the other hand, the statistical reliability of the estimates strongly depends on the available temporal data points. That is, the larger is the number of available data points, the better is the ability of the FC estimator to capture the underlying connection mechanism. This situation is further exacerbated in the case of multivariate and non-linear estimators, which typically require more data points to give reliable estimates [70,71]. Standard solutions consist in concatenating the temporal windows associated with multiple repetitions of the same experimental task or averaging the FC estimates obtained in each repetition [72]. Methods based on multi-window spectrum estimation can be also used [73,74]. They allow the analysis of short data segments by using smoothing over orthogonal windows and they can be defined in the Fourier [75] and Wavelet domain [76]. Another elegant approach to estimate time-varying FC would consist in the use of methods formally designed to deal with non-stationary signals, such as detrended fluctuation analysis [77] or wavelet decomposition [78]. Among others, time-frequency methods such as *wavelet coherence* [58,79] and *adaptive partial directed coherence* [52,80] represent particularly appealing solutions.

More in general, the development of FC methods able to capture time-varying interactions is a fertile research field. For instance, tracking algorithms of brain correlation dynamics have been recently exploited [81], also considering low-rank subspaces [82]. Other approaches are based on model assumptions on the nature of signals [83], time-varying autoregressive models and variation of standard connectivity estimators [47,84].

## 2.2. Choosing the best FC estimator

We reported some of the most common FC estimators and their associated ability to solve one or more criticalities. It is important to notice that in general none of them is able to simultaneously solve all the raised issues. While it may be expected that applying all the possible methods would lead to consistent results, this approach lacks a precise rationale because different estimators intrinsically capture different signal properties and address different methodological questions (eg, causality *versus* synchronization). Instead, the choice of the "best" estimator mainly depends on the specific scientific question [85]. If the scientific hypothesis that guides the analysis is clear, the choice of the estimator should be a natural consequence. For example, if the goal of the study is to determine information flows between two brain areas, a directed estimator should be used in a bivariate framework. In this scenario, under the assumption of linear dynamics, linear methods such as *Granger causality* should be used, otherwise nonlinear estimators such as *transfer entropy* should be preferred.

Particular attention should be paid when studying brain signals with rich frequency dynamics. The use of estimators defined in the frequency domain is well-suited if the goal is to determine FC at specific frequency bands. The frequency transformation implemented by these estimators is typically obtained either via parametric techniques, such as autoregressive models, or non-parametric techniques such as Fourier or Hilbert transformations. In the case of temporal-domain FC estimators, it is still possible to derive estimates in the frequency domain by pre-bandpassing the signals, e.g. *phase-locking value*.

Another element involved in the choice of the estimator is the temporal resolution of the neuroimaging technique. In fact, EEG and MEG signals are characterized by high temporal dynamics, in the order of milliseconds, while fMRI data exhibit low temporal dynamics, in the order of seconds. Thus, EEG and MEG signals can exhibit signal changes in a very broad frequency range, from infraslow (<1 Hz) to ultra-fast (>100 Hz) dynamics depending on the task and on the presence of pathological conditions [86–88]. For this reason, frequency-domain methods are more appropriate with EEG/MEG signals as they allow to isolate FC in specific frequency bands of interest. On the contrary, time-domain methods, such as Pearson correlation and partial correlation [89], can be more appropriate with fMRI data, where the available frequency range is rather limited (i.e. < 1 Hz) [90].

## 2.3. From connectivity to networks

After computing FC for each pair of signals, the corresponding values can be collected in the so-called connectivity matrix $A$, i.e. a $N \times N$ matrix, where $N$ is the number of nodes (sensors, ROIs, …) and the entry $a_{ij}$ contains the FC value for the connection, or edge, between the nodes $i$ and $j$.

Diagonal elements $a_{ii}$ correspond to FC of a node with itself. Because their interpretation is not trivial, the main diagonal of the connectivity matrix is typically set to null values. In addition, in presence of directed FC the direction of the connection must be specified to correctly read the connectivity matrix. In fact, while for undirected FC there is a symmetric relation between the elements of the connectivity matrix (i.e., $a_{ij} = a_{ji}$), for directed FC the relation becomes asymmetric (i.e., $a_{ij} \neq a_{ji}$).

The values contained in the connectivity matrix depend on the nature of the employed FC estimator. While the majority of the methods give normalized values within the $[0,1]$ interval, there might be in general different ranges or scales. In these situations, it is often preferable to transform the data, taking into account the nature of the FC estimator, so to rescale them within the normative interval. For example, Pearson correlation gives values that span the interval $-1 \leq a_{ij} \leq 1$, i.e. from perfect anticorrelation (anti-phase) to perfect correlation (in-phase). However, since it might be difficult to interpret the negative values from a neurophysiological perspective (i.e., true anti-phase behavior or simple delayed interaction), a common procedure is to consider the absolute values in the corresponding connectivity matrix and interpret their magnitude as general correlation.

Statistical approaches based on known properties of the estimators or on data surrogates can be eventually used to remove non-significant FC values [3].

## 3. Network science to study functional connectivity matrices

Together, nodes and edges form a new type of networked data that cannot be studied with standard tools, but needs appropriate techniques from *network science*, i.e. a modern field that draws on graph theory, statistical mechanics, data mining and inferential modeling [91,92]. Network science allows to analyze *complex systems* at different spatial scales – from molecular biology to social sciences – and to quantify organizational mechanisms by extracting indices that characterize specific topological properties [10,93].

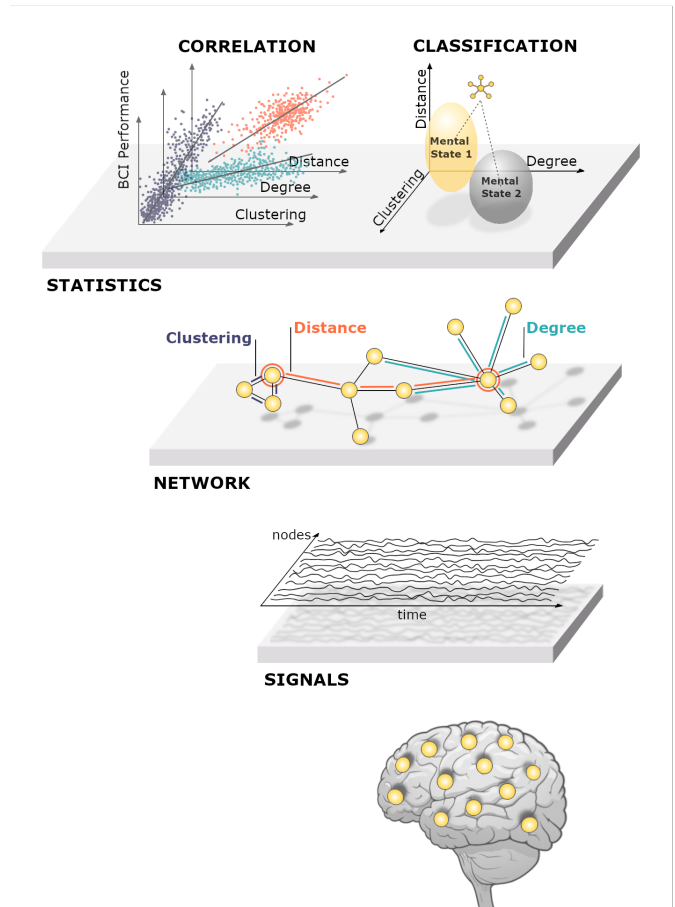

**Figure 1 - Principal scheme of a network-based brain-computer interface.** From bottom to top: brain networks are reconstructed by computing functional connectivity between remote brain signals (Section 2). The resulting connectivity networks are characterized by means of graph theoretic metrics, which extract summary indices quantifying different topological properties (Section 3). These values correspond to specific network properties that can be used to identify predictors of BCI performance (Section 4) as well classify different BCI mental states (Section 5).

In this framework complex networks are modeled as graphs, i.e. mathematical objects defined by nodes and edges [93]. After being constructed, the resulting brain network corresponds to a weighted graph whose edges code for the magnitude of the FC between different nodes. Common courses in brain network analysis typically use thresholding procedures to filter the raw networks by retaining, and eventually binarizing, a certain percentage of the available links. These procedures typically result in sparse networks with a relatively low connection density (**Box 1**). Despite the consequent information loss, thresholding is often adopted to

---

**Box 1 - Basic characteristics of graphs**

**Density:** ratio of actual number of edges and the number of total possible edges in the network. Brain networks tend to be relatively sparse (i.e. density < 50%) [104], although there is a high variability due to thresholding procedures.

**Walk, cycle and path:** a walk is a sequence of successive edges which joins a sequence of nodes. A cycle is a closed walk where the first and last nodes coincide. A path is a walk in which all edges and nodes are distinct. A graph is said to be connected if there exists a path between any possible node pair.

**Distance:** length of the shortest path between two nodes. In weighted graphs the shortest path is the one that minimizes the sum of the edge weights along the path. In brain networks, weights should be inverted when computing distances as the highest weights correspond to the strongest, most reliable, links [3,105].

---

mitigate the uncertainty of the estimated weakest edges, reduce the false positives, and facilitate the interpretation of the inferred network topology [94–96].

The simplest way to proceed is to fix a threshold on the number of strongest links to retain or on the FC value. However, these approaches are parametric and researchers are often required to repeat the analysis for a broad range of different thresholds and eventually select the one belonging to an interval for which results remain relatively stable.

Since these approaches might be considered suboptimal, they can be alternatively replaced by theoretically-grounded nonparametric methods based on different criteria including statistical contrasts with data surrogates [97,98], topological optimization [99–101,94] and population-based consensus [102,103].

After thresholding, network properties can be extracted from the resulting sparse networks, which can be weighted or unweighted depending on whether the remaining weights are maintained or binarized. For the sake of simplicity, we will describe in the following graph theoretic metrics in the case of undirected and unweighted networks, mentioning how they can be extended in the general cases.

## 3.1. Network metrics

In this section, starting from general notions, we present the main network metrics to quantify local-, meso-, and global-scale topological properties of brain networks or graphs. Local scale properties are at the level of a single node, and quantify its importance in the network according to different criteria.

Meso-scale properties refer to grouping of nodes based on distinctive interaction patterns. Global-scale properties characterize the network as a whole and represent a summary index.

As a reminder, we refer to $A$ as to the connectivity or adjacency matrix of the filtered brain network containing $N$ nodes and $L$ links, or edges.

### 3.1.1. Local-scale properties

#### 3.1.1.1. Degree

The most intuitive metric for a node is the so-called node degree which counts the number of connections with the rest of the network. For binary, undirected networks the degree of node $i$ can be computed as

$$k(i) = \sum_{j=1, j \neq i}^{N} A_{ij} \qquad 1)$$

The analog of node degree in weighted networks is known as node strength, which simply sums the weights of the connections of node $i$ to the rest of the network. In the case of directed graphs, it's possible to both count the number of incoming edges of node $i$, and the number of outgoing edges considering the sum of the rows or columns of $A$.

The node degrees are generally used to identify the most connected nodes in the graph that hold a large part of the overall system's connectivity and therefore represent candidate hubs of the brain network.

#### 3.1.1.2. Betweenness

Apart from the node degree, there are in general several ways in which a node can be considered central or important in a network. Betweenness centrality measures the extent to which a node lies "between" other pairs of nodes by considering the proportion of shortest paths (**Box 1**) in the network passing through it [106,107]. In practice, the betweenness centrality of a node $i$ reads as

$$\mathcal{C}_B(i) = \frac{1}{(N-1)(N-2)} \sum_{h=1, h \neq j}^{N} \sum_{j=1, j \neq i}^{N} \frac{\sigma_{hj}(i)}{\sigma_{hj}} \quad 2)$$

where $\sigma_{hj}(i)$ is the number of shortest paths between nodes $h$ and $j$ that pass through $i$, $\sigma_{hj}$ is the number of shortest paths between nodes $h$ and $j$. Betweenness centrality can be computed in the same way for weighted and directed networks, i.e. calculating the shortest paths following the direction of the edges.

Assuming that information flow along shortest paths, the betweenness centrality can be used to identify those nodes which are crucial for the information transfer between topologically distant brain regions.

### 3.1.1.3. Communicability

Differently from betweenness centrality, communicability takes into account the contribution of all possible walks between node pairs [108]. By doing so, communicability reflects a network's capacity for parallel information transfer.

Formally, the communicability of a node $i$ is given by

$$\mathcal{C}_C(i) = \sum_{j=1}^{N} [e^A]_{ij} \qquad 3)$$

where $e^A$ denotes the matrix exponential of the matrix $A$ that takes into account for each pair of nodes the total number of walks between them [109]. Communicability in weighted networks can be computed by normalizing the connectivity matrix with appropriate transformations [110]. In the case of directed networks, heuristic approaches can be used to identify all the possible paths of a specified maximum length [111].

Communicability can be particularly suitable for identifying brain areas that are central for the diffusion of information across the network [110,112].

### 3.1.1.4. Eigenvector

The eigenvector centrality of a node is a metric which considers the importance of its neighbors, i.e. the nodes directly connected, or adjacent, to it. Hence, it can be thought as being equivalent to the summed centrality of its neighbors [113]. The eigenvector centrality of a node $i$ is obtained by computing graph spectrum and reads as

$$\mathcal{C}_E(i) = \frac{1}{\lambda} \sum_{j=1}^{N} A_{ij} v_j \qquad 4)$$

where $\lambda$ is the largest eigenvalue of $A$ and $v$ is the associated leading eigenvector. Eigenvector centrality can be extended to weighted networks, subject to certain conditions [93,114]. In this case, $A$ must be positive definite and this condition might not be satisfied for correlation-based networks which also contain negative entries. One solution is to remap edge weights to a positive range, by taking for instance the absolute value of the correlation coefficients. In directed networks, the adjacency matrix $A$ is asymmetric and there are two leading eigenvectors, which can be therefore used to isolate the contribution of either incoming or outgoing edges.

Eigenvector centrality can be used to identify brain areas which do not necessarily have a high number of links, but that are connected to other central regions [115].

### 3.1.2. Meso-scale properties

### 3.1.2.1. Motifs

Network motifs are subgraphs that repeat themselves in a network. Each of these subgraphs, defined by a particular pattern of interactions between nodes, often reflects a mode in which particular functions are realized by the network.

The motif detection can be done under various paradigms including exact counting, sampling, and pattern growth methods [116]. After calculating the frequency $F$ -as the number of occurrences- of a subgraph $G$ the assessment of its significance is given by

$$Z(G) = \frac{F(G) - \mu(G)}{\sigma(G)} \qquad 5)$$

Where $\mu$ and $\sigma$ indicate respectively the mean and standard deviation of the frequency of the subgraphs in an ensemble of random networks corresponding to a null-model associated to the empirical network (see next subsection). The resulting Z-score indicates if the motif $G$ is occurring either more or less than expected by chance. While motif detection naturally applies to binary networks, the extension to weighted ones can be achieved by replacing the motif occurrence with its intensity [117].

Motifs represent the basic building blocks of a network and may provide a deep insight into the brain network's functional abilities [118,119], albeit their detection is computationally challenging as the number of nodes becomes higher than six [120].

### 3.1.2.2. Communities and modularity

Communities, or modules, are often defined in terms of network partitions where each node is assigned to one and only one module. Community detection structure is not trivial and many algorithms to identify community structures are available. For instance, they may be based on hierarchical clustering, spectral embedding, statistical inference, and more recently machine learning approaches [121,122].

The quality of the identified partition can be measured by the so-called modularity index

$$Q = \frac{1}{2L} \sum_{i=1}^{N} \sum_{j=1}^{N} (A_{ij} - R_{ij}) \delta(m_i, m_j) \qquad 6)$$

where $R_{ij}$ is the probability to observe an edge as expected by chance and the Kronecker delta $\delta(m_i, m_j)$ equals one if nodes $i$ and $j$ belong to the same module (i.e., $m_i = m_j$) and zero otherwise. When $Q$ is positive, the network tends to have high intramodule connectivity and low intermodule connectivity; when $Q$ is less than or equal to zero then the network lacks a modular structure. The above equation can be extended to the analysis of weighted [114], and directed networks [123].

In brain networks, topological modules tend to be spatially localized, and they typically include cortical areas that are known to be specialized for visual, auditory, and motor functions [124].

### 3.1.2.3. Core-periphery structure

Core-periphery is a peculiar partition of the network consisting of a group of tightly connected nodes (i.e. the core), and a group made by the remaining weakly connected nodes (i.e. the periphery) [125]. Identifying the core of a network can be achieved through methods optimizing a fitness function or via statistical null models [126]. These methods rely on subjective fine-tuning of one or more free parameters and tend to be relatively complex with consequent scalability issues.

Here we report an alternative method that only requires the degree sequence and no prior knowledge on the network [127]. The basic idea is to separate the nodes in two groups based on their rank, as determined by their node centrality (e.g. the degree). The optimal separating rank position is then given by

$$r^* = \underset{r}{\mathrm{argmax}}(k_r^+) \qquad 7)$$

where $k_r^+$ is the number of links that a node of rank $r$ shares with nodes of higher rank. This method has the advantage of being fast, highly scalable and it can be readily applied to weighted and directed networks.

In brain networks, core-periphery organization is thought to emerge as a cost-effective solution for the integration of distributed regions in the periphery [128]. A related concept is that of rich-club behavior, where the brain network hubs tend to be mutually interconnected [129] .

### 3.1.3. Global-scale properties

#### 3.1.3.1. Characteristic path length and global-efficiency

The characteristic path length is a scalar that measures the global tendency of the nodes in the network to integrate and exchange information. Assuming that the information flows through the shortest paths, the characteristic path length is given by [130]

$$P = \frac{1}{N(N-1)} \sum_{i=1, i \neq j}^{N} d_{ij} \qquad 8)$$

where $d_{ij}$ is the distance between nodes $i$ and $j$. Because the distance between two nodes that are not connected through any path is equal to infinity, $P$ is ill-defined for disconnected networks.

To overcome this issue, the *efficiency* between two nodes as the reciprocal of their distance was introduced. With this measure the contribution of two disconnected nods becomes zero. Hence, the global-efficiency of a network is a normalized scalar given by [5]

$$E_{glob} = \frac{1}{N(N-1)} \sum_{i=1, i \neq j}^{N} \frac{1}{d_{ij}} \qquad 9)$$

Both $P$ and $E_{glob}$ can be easily applied to directed and weighted networks taking into account the appropriate way to compute the distance (**Box 1**). Characteristic path length and global-efficiency represent two of the most widely used measures of integration in brain networks because of the simplicity of their interpretation [2].

An average short distance between the nodes may constitute a biological mechanism to minimize the energetic cost associated with long-range connectivity, and could provide more efficient and less noisy information transfer [1,131].

#### 3.1.3.2. Clustering coefficient and local-efficiency

Clustering is an important feature in complex networks that measures the extent to which nodes' neighbors are mutually interconnected. Strongly related to the presence of triangles in the network (i.e. triads of nodes fully connected), the clustering coefficient is a normalized scalar given by

$$C = \frac{1}{N} \sum_{i=1}^{N} \frac{2l_i}{k_i(k_i - 1)} \qquad 10)$$

where $l_i$ is the number of links between the neighbors of node $i$ and $k_i$ its node degree. The extension to weighted and directed networks was proposed in [117,132].

Alternatively, the overall tendency of a network to form a clustered group of nodes can be obtained in terms of network global-efficiency. The so-called local-efficiency is given by averaging the global-efficiencies of the network's subgraphs

$$E_{loc} = \frac{1}{N} \sum_{i=1}^{N} E_{glob}(G_i) \qquad 11)$$

where $G_i$ denotes the subgraph comprising all nodes that are immediate neighbors of the $i^{th}$ node. In brain networks, the clustering coefficient and local-efficiency are often interpreted as a measure of functional segregation or specialization [133].

| Network metric | | Computational complexity | |
|---|---|---|---|
| | | Unweighted | Weighted |
| *Local-scale* | Degree | $O(L)$ | $O(L)$ |
| | Betweenness | $O\big(N(N+L)\big)$ | $O\big(N(L+N\log N)\big)$ |
| | Communicability | $O(N^3)$ | $O(N^3)$ |
| | Eigenvector | $O(N^2)$ | $O(N^2)$ |
| *Meso-scale* | Motifs | $O(Ng)$ | - |
| | Communities | $O(N\log N)$ | $O(N\log N)$ |
| | Modularity | $O(L)$ | $O(L)$ |
| | Core-periphery | $O(L+N\log N)$ | - |
| *Global-scale* | Characteristic path length | $O\big(N(N+L)\big)$ | $O\big(N(N+L)\big)$ |
| | Global-efficiency | $O\big(N(N+L)\big)$ | $O\big(N(N+L)\big)$ |
| | Clustering coefficient | $O(L\langle k^2\rangle/\langle k\rangle)$ | $O(L\langle k^2\rangle/\langle k\rangle)$ |
| | Local-efficiency | $O\big(N(\langle k^2\rangle-\langle k\rangle)\big)$ | $O\big(N(\langle k^2\rangle-\langle k\rangle)\big)$ |

**Table 2 - Computational complexity of network metrics**. $N$ = number of nodes; $L$ = number of links; $g$ = size of the motif; $\langle k\rangle$ = average node degree; $\langle k^2\rangle$ = node degree variance.

Together with distance-based metrics ($P$ and $E_{glob}$), clustering metrics are used to quantify the small-world properties of a network, i.e. the tendency to optimize simultaneously integration and segregation of information [130]. Because the strong parallel with a plausible model of neural functioning these metrics are the most widely used in the field of network neuroscience [134].

In practice, the small-world propensity can be computed by normalizing the values of the empirical network with those obtained from network surrogates, such as random graphs [135]. Then, a small-world index can be obtained, for example, by combining the normalized $P$ and $C$ values

$$\omega = \frac{C}{\mu(C_{rand})}\frac{\mu(P_{rand})}{P} \qquad 12)$$

where $P_{rand}$ and $C_{rand}$ are vectors containing the values obtained for the network surrogates. Notably, other types of small-world indexes can be obtained by opportunely substituting $P$ and $C$, with $E_{glob}$ and $E_{loc}$ [136], or by adopting normalization with other types of network surrogates [137].

We report in **Table 2** the time complexity of the above metrics for unweighted and weighted networks [134].

### 3.2. Normalizing network metrics

Most measures of network organization scale with the number of nodes and edges in a graph. Thus, to compare the values of the metrics extracted from different size and connection densities, it is often necessary to account for basic properties of the underlying network. As mentioned before, normalization with respect to null, or reference, models provides a practical benchmark to determine the extent to which a network property deviates from what would be expected by chance and to compare network properties across different conditions [2,138,94].

Generating reference networks that match all properties of an actual network except for the one that has to be normalized is difficult in practice, since most properties are interrelated. It is therefore usual to match only basic properties, such as network size, connection density, and degree distribution. This kind of null network is typically obtained using randomization strategies, where the actual network is randomly rewired according to a set of rules. In particular, the rewiring may be performed either preserving the degree distribution or not, the former being a more conservative choice [139].

Because the rewiring process is stochastic, a certain number of network samples - typically higher than 100 - should be generated in order to constitute an ensemble of reference networks with similar characteristics.

The normalized value of a metric can then be computed as the ratio of the value measured on the observed network ($M_{obs}$) and the mean obtained from the randomized network ensemble

$$M' = \frac{M_{obs}}{\mu(M_{rand})} \qquad 13)$$

While the ratio is the preferred way to normalize network metrics, Z-scores procedures can be used as well (Eq. 5). Notably, rewiring procedures that preserve the degree distribution have been extended to weighted and signed networks [140]. While generating purely random network ensembles is the most intuitive way of normalizing, alternative strategies that generate more complex null models might be adopted, too (**Box 2**).

## 3.3. Advanced network approaches

The previous paragraphs introduced some of the well-established graph metrics used in network neuroscience that might be particularly relevant to BCI applications. Nonetheless, the field of network science is quickly advancing and new research directions are currently in development to address the open challenges.

First, the abovementioned graph metrics have been mainly conceived as topological descriptors of static networks, whose links do not change in time. This is an oversimplification of the real phenomena as brain networks are intrinsically dynamic and functional connectivity can change across multiple time scales (eg, within and between BCI sessions). Hence, time must be formally considered as a part of the network problem and not merely as a repeated measure [141]. In neuroscience, many network metrics have been rethought temporally by considering the nature of time-respecting paths [142] and the persistence of specific motifs [119] and modules [143]. The theoretical development of temporal networks appears therefore particularly relevant for future BCI-related studies.

Second, the characteristics of the brain network strongly depend on the neuroimaging technique (i.e., the nodes) and on the type of functional connectivity estimator used (i.e., the edges). That means that multiple brain networks simultaneously characterize the same subject. Multilayer networks have been recently introduced to provide theoretically grounded metrics integrating the available information from multiple sources [144,145]. In multilayer brain networks, different types of connectivity are represented on different layers (eg, neuroimaging modality [128] and frequency bands [146,147]) and connectivity can span both within and between layers (eg, cross-frequency coupling [148]). Notably, multilayer network metrics are able to extract higher-order information that cannot be obtained by simply aggregating connectivity across layers. Therefore, this innovative framework for integrating different connectivity levels might be particularly useful for the development of multimodal BCI systems [149].

Together with the descriptive nature of the network metrics (which are intrinsically data-driven) the development of network models could greatly advance the study of brain networks in BCI by providing complementary statistical information. Since brain networks, as in other real networks, are typically inferred from experimental data their edges are subject to statistical uncertainty. Stochastic network models based on spatial, topological or Bayesian rules have been recently introduced to tackle those aspects and obtain a more robust understanding of the organizational properties of complex brain networks [150–152]. Finally, approaches based on network controllability [153] could be used in brain networks to identify the driver nodes that could be experimentally targeted by BCI feedback to elicit specific mental states or behaviors [154].

## 4. Network properties underlying BCI motor tasks

BCIs involve a complex mixture of cognitive processes not necessarily directly linked with the targeted task [155,156]. Among them are attention and task engagement [157], working memory and decision-making [158–161], but also error-potential have been shown to occur during BCI tasks [162–164]. These higher-order cognitive processes result from interactions between different areas that engender brain network reorganization. Here, we will specifically focus on the network changes underlying motor (executed and imagined) performance, which is largely studied in the literature and directly associated with one of the most used BCI paradigms.

### 4.1. Short-term dynamic network changes during motor tasks

Performing motor imagery-based BCI experiments consists of the voluntary modulation of $\mu\beta$ activity to control an object [165]. The analysis of event-related desynchronization and event-related synchronization enables the detection of mental states [18,166–169]. Notably, motor imagery (MI) and execution (ME) tasks have been shown to share similar characteristics such as the spatial and frequency localization of the evoked brain activity [170–172].

Meta-analyses, mainly based on fMRI and PET studies, recently revealed a group of regions involved during ME [173] and MI [174], including premotor area, primary sensorimotor area, supplementary motor area, posterior parietal lobe. Notably, Hardwick et al. [175], made a comparison between imagery, observation and execution. They identified two main clusters involved in both MI and ME: bilateral cortical sensorimotor and premotor clusters. They also performed contrast analyses to elicit regions more consistently involved in MI than in ME. It appeared that MI tends to recruit more often premotor regions and left inferior and superior parietal cortex. These results seem to be corroborated by studies performed from a network perspective. By using betweenness centrality, Xu et al. [176] showed that in ME the most important region lies in the supplementary motor area (SMA) whereas during MI the most central area was located in the right premotor area (rPMA). In the case of ME, it would suggest that SMA could enable an efficient communication between brain areas, especially motor ones [177,178] during sequential execution. In the case of MI, PMA could integrate both sensorimotor information from motor areas (e.g. SMA) and spatial information of movements from regions such as posterior parietal lobe to enable motor planning [177,179,180].

Complementary to the previous studies, another approach consists of studying time-varying network properties while performing tasks [119,181,182]. In the specific case of motor tasks, a work based on the use of time-varying partial direct coherence (PDC) revealed that the cingulate motor areas could be seen as a hub of outgoing flows during dorsal flexions of the right foot [182]. Based on experiments performed with five subjects via a 64-EEG channel system, the authors observed changes of network patterns at different stages of the task. The preparation of the movement presented a high level of efficiency, associated with an increase of clustering coefficient and a reduction of the characteristic path length. During the movement, strong functional links between the cingulate motor and the supplementary motor areas were obtained but also a lower network efficiency at the global level. These results illustrate the existence of a dynamic network reorganization process during the preparation and execution of a simple motor task.

## 4.2. Long-term longitudinal network changes in learning

Understanding how we learn to use a BCI is crucial to adapt to individual variability and improve performance. Learning is a complex phenomenon that can be categorized in different types such as instructed (supervised [183] or reinforced [184]) or unsupervised [185], explicit or implicit [186].

Regardless of the type of learning, it is characterized by changes in brain associations from microscale, with the synapse strengthening for example, to macroscale levels, including changes of functional brain connectivity. In this section, we present some of the recent studies using network science approaches to characterize large-scale neural processes of human learning at the macroscale [187,188].

### 4.2.1. Motor learning

In the past years, studies focusing on functional connectivity demonstrated changes induced by motor skill learning. Comparisons made before and after a locomotor attention training revealed an alteration of the connectivity in the sensorimotor areas potentially modulated by focusing attention on the movements involved in ambulation [189]. Sensorimotor adaptation tasks involve notably prefrontal cortex, premotor and primary motor and parietal cortices [190] and once acquired, motor skills are encoded in fronto-parietal networks [191]. However, little is known about its evolution through training.

In Taubert et al. [192], fourteen healthy subjects performed a dynamic balance task once a week during six consecutive weeks. They underwent four fMRI scans: before the first, the third, the fifth sessions and one week after the training program. The authors observed an increased fronto-parietal network connectivity in one week. Training sessions progressively modulated these modifications. Changes induced by motor imagery learning have been observed, notably in resting-state functional connectivity of the default mode network (DMN) [193]. These results prove that motor learning relies on areas beyond those directly involved during the task performance and illustrate the need to study how communication between brain regions evolves during the training.

From a network perspective, a large number of metrics characterizing the topological properties have been considered to capture the motor acquisition process. Heitger et al. [194] showed that the motor performance improvement of a complex bimanual pattern was associated with an increase of clustering coefficient and a shorter communication distance. However, it should be considered that the latter one was possibly influenced by the reported higher connection density and strength.

Network modularity has been used as a marker in the case of age-related changes [195] but also in the case of induced brain plasticity [196]. Therefore, it seems particularly of interest in the study of learning process as it captures changes in the modular organization of the brain [197]. In the specific case of motor skill acquisition based on the practice of finger-movement sequences over six weeks, the use of modularity revealed that learning induced an autonomy of sensorimotor and visual systems and individual differences in amount of learning could be predicted by the release of cognitive control hubs in frontal and cingulate cortices [198].

Based on the temporal extension of network modularity, Bassett et al. [197] defined the "flexibility" as the number of times a node changes its module allegiances between two consecutive time steps. This measure was used to study the evolution of brain network properties during a motor learning task. Twenty-five healthy subjects were instructed to generate responses to a visually cued sequence by using the four fingers of their non-dominant hand. They participated in three training sessions in a five-day period, performed inside the fMRI. The flexibility predicted the relative learning rate, particularly in frontal, pre-supplementary motor, posterior parietal and occipital cortices [197].

### 4.2.2. Neurofeedback and BCI learning

To master closed-loop systems such as neurofeedback (NFB) or brain-computer interfaces, several training sessions are typically needed. Recent studies suggest that the involved learning process is analogous to cognitive or motor skill acquisition [199]. NFB could induce behavioral modifications and neural changes within trained brain circuits that last months after training [200]. At microscale, changes at the neuronal level have been observed and simulated during BCI learning [201]. At larger spatial scales, the recruitment of areas beyond those targeted by BCI has been observed during the skill acquisition [202,203]. For example, the decrease of the global-efficiency in the higher-beta band indicated the involvement of a distributed network of brain areas during MI-based BCI training [204]. These findings motivated a deeper understanding of the brain network reorganization, at the macroscale, underlying the BCI/NFB learning process.

In a recent study, Corsi et al. [205] studied how the brain network reorganizes during a MI-based BCI training. Twenty healthy, and BCI-naïve, subjects followed a four-session training over two weeks. The BCI task consisted of a standard 1D two-target task [206]. To hit the up-target, the subjects had to perform a sustained MI of right-hand grasping and to hit the down-target they remained at rest. MEG and EEG signals were simultaneously recorded during the sessions.

Results obtained from the relative node strength showed a progressive reduction of integration among, primary visual areas and associative regions, within the α and β frequency ranges. This metric could also predict the learning rate more specifically in the anterior part of the cingulate gyrus and the orbital part of the inferior frontal gyrus, both known to be involved in human learning [207], and the fronto-marginal gyrus and the superior parietal lobule, which is associated with learning and motor imagery tasks [208,209]. To fully take advantage of the behavioral and MEG information to predict learning, a multimodal network approach has been adopted by Stiso et al. [154]. The authors used a non-negative matrix factorization to identify regularized, covarying subgraphs of functional connectivity to estimate their similarity to BCI performance and detect the associated time-varying expression. From their observations, they deduced a model tested via the network control theory in which specific subgraphs support learning via a modulation of brain activity in areas associated with sustained attention.

Despite the promising evidence, brain network reorganization needs to be further investigated to better understand learning mechanisms underlying the use of BCI devices and enhance the usability in clinical applications [21,203].

## 4.3. Clinical applications: the case of stroke

It is well known that neurological or psychiatric disorders lead to changes in terms of communication between brain regions [8]. For example, connectivity between high-degree hub nodes has been observed in schizophrenia [210] and comatose patients [211]. Decreased global- and local-efficiencies has been reported in Parkinson disease [212], while modifications of the core-periphery structure [213] and a loss of inter-frequency hubs has been found in Alzheimer disease [147]. In the case of attention-deficit/hyperactivity disorder in children the increase of local-efficiency and lower global efficiency suggested a disorder-related tendency toward regular organization [214]. In addition, modifications in nodal properties have been observed in both children and adults in the attention, sensorimotor and DMN [215] and striatum [214,216,217].

Brain network changes in stroke patients are particularly relevant for BCI clinical applications and neurofeedback rehabilitation strategies. Recent studies showed that stroke recovery is accompanied by an increased smallworldness, which supports increased efficiency in information processing [218,219]. Laney et al. [220] performed a study with ten stroke patients that participated in six-weeks training sessions dedicated to improve voluntary motor control. fMRI data were collected, before and after training, while patients performed an auditory-cued grasp and release task of the affected hand. Finger extensions were assisted by an MRI compatible exoskeleton. Two opposite effects were observed: an increased node closeness-centrality [10] with sensorimotor and cerebellum networks and a decreased closeness-centrality in the DMN and right frontal-parietal components. The authors associated the former to an improved within-network communication and the latter to a reduced dependence on cognition as motor skill enhanced [220]. In another study [221], authors aimed to characterize the brain network reorganization after stroke in the chronic stage in a group of twenty patients. Brain networks were constructed by estimating wavelet correlation from fMRI signals. They showed an overall reduction of connectivity in the hubs of the contralesional hemisphere as compared to healthy controls. Most of these studies are based on a static representation of the brain plasticity and partially inform on the individual ability of stroke patients to recover motor or cognitive functions. Recently, an approach based on temporal network models that aimed at tackling these issues indicated that both the formation of clustering connections within the affected hemisphere and interhemispheric links enabled to characterize the longitudinal network reorganization from the subacute to the chronic stage [222]. These mechanisms could predict the chronic language and visual outcome respectively in patients with subcortical and cortical lesions.

Motor imagery has been proved to be a valuable tool in the study of upper-limb recovery after stroke [223]. It enabled observations of changes in ipsilesional intrahemispheric connectivity [224] but also modifications in connectivity in prefrontal areas, and correlations between node strengths and motor outcome [225]. Within the beta frequency band, performing a MI task of the affected hand induced lower small-worldness and local-efficiency compared to the MI of the unaffected hand [136]. Based on previous observations in

**Box 2 - Network generative models**

**Random** networks are generated with the Erdös-Rényi (ER) model. They are constructed by fixing a parameter $p$ which fixes the probability to have a link between two randomly selected nodes in the graph. By construction, $p$ coincides with the connection density of the resulting networks. In general, ER networks do not exhibit any particular structure but typically low characteristic path lengths [228].

**Small-world** networks are generated with the Watts-Strogatz (WS) model. Starting from a ring lattice graph, where each node is connected to its first $k$ neighbors, WS networks are generated by rewiring the links with a probability $p_{WS}$, i.e. the model parameter. With relatively low values of $p_{WS}$, the resulting networks exhibit both high clustering coefficient and low characteristic path length. This is a feature observed in many real-world interconnected systems and it optimizes both segregation and integration of information [130].

**Scale-free** networks are generated with the Barabási-Albert (BA) model. Its construction starts with $m_0$ nodes. Then, new nodes are iteratively added with $m$ ($\leq m_0$) links that connect them to existing nodes with a probability $p_{BA}$ proportional to their node degree. As a result of such preferential attachment rule, BA networks show highly heterogeneous node degrees, few strongly connected hubs as well as low characteristic path length and null clustering coefficient. These features have been found in many real networks as a sign of resilience [229].

**Modular** networks are generated with the stochastic block model (SBM). This model partitions the nodes in $M$ groups of arbitrary size. Then it assigns edges between nodes with a probability that fixes the expected connection density within- ($p_{intra}$) and between-groups ($p_{inter}$). By construction, SBM networks have high modularity values as well as typical small-world properties [230].

resting-state [226], a recent double-blind study revealed that node strength, computed from the ipsilesional primary motor cortex in the alpha band, could be a target for a motor-imagery-based neurofeedback and lead to significant improvement on motor performance [227].

# 5. Network features for improving BCI performance

The use of network approaches in BCI is a relatively young and unexplored area, yet, the existing publications show encouraging results. In this section, we first provide a proof-of-concept on simulated data to illustrate the theoretical benefit of using network metrics from a classification perspective. Then, we present some of the recent classification results obtained with neuroimaging data during real BCI experiments.

In current settings, different mental strategies are used to control the MI-based BCI. The resulting brain states are translated into features that need to be properly recognized by the classifier. To reproduce this scenario, we associated different brain states with networks having distinct topological properties. Specifically, we generated synthetic networks exhibiting four different topologies, or classes, which have been extensively reported in neuroscience, i.e. small-world, modular, scale-free and random networks [8,231]. These networks were generated with the models described in **Box 2**. We fixed the same number of nodes ($N = 100$) and links ($L = 600$) for all of them. The specific model parameters values were: $p = 0.12$ for random networks; $p_{WS} = 0.1$ for small-world; $m_0 = m = 6$ for scale-free; and finally $M = 4$ of equal size for modular ones, with $p_{intra} = 0.46$ and $p_{inter} = 0.01$. These networks qualitatively exhibit disparate properties in terms of integration, segregation and heterogeneity of information (**Figure 2A**). To quantify these differences, we computed four relevant network metrics, i.e. global-efficiency, local-efficiency, modularity and degree-variance. In order to sample the distribution of these properties across models, we generated a large ensemble of 1000 networks per class (**Figure 2B**).

We then evaluated the performance of network metrics in discriminating the four classes as compared to the use of the entire connectivity matrix. We specifically tested 2-classes and 4-classes scenarios according to the typical number of mental states used in BCIs. To reproduce the fact that nodes might not correspond exactly to the same brain areas across different subjects - because of a natural individual spatial and functional variability [232] - we further performed a random permutation of the node labels. Notably, this procedure did not alter the intrinsic topology of the generated networks.

Classification accuracies were finally obtained from a repeated random sub-sampling validation with 100 random balanced-splits. Specifically, the training set consisted of 80% of all the networks, while 20% of the networks were used as testing set. Results showed that when we applied node permutation, the average classification accuracy of connectivity matrices progressively decreased down to chance levels, while network metrics always exhibited a perfect classification. More precisely, from 50% of node permutation, the accuracy obtained by using connectivity matrices was significantly lower than network metrics (Wilcoxon test, $p < 1.6 \times 10^{-10}$, Bonferroni corrected for multiple comparisons) (**Figure 2C**).

All network analysis and classification have been performed with the freely available NetworkX and Scikit-learn packages in Phyton.

Taken together, these results indicated the theoretical benefit of including network metrics into the classification of BCI-related mental states. The development of sophisticated machine learning techniques, which operate on the entire connectivity matrices [233], could lead to similar performance in the next future, too. Finally, it is important to mention that the advantage of network metrics also lies in their relatively

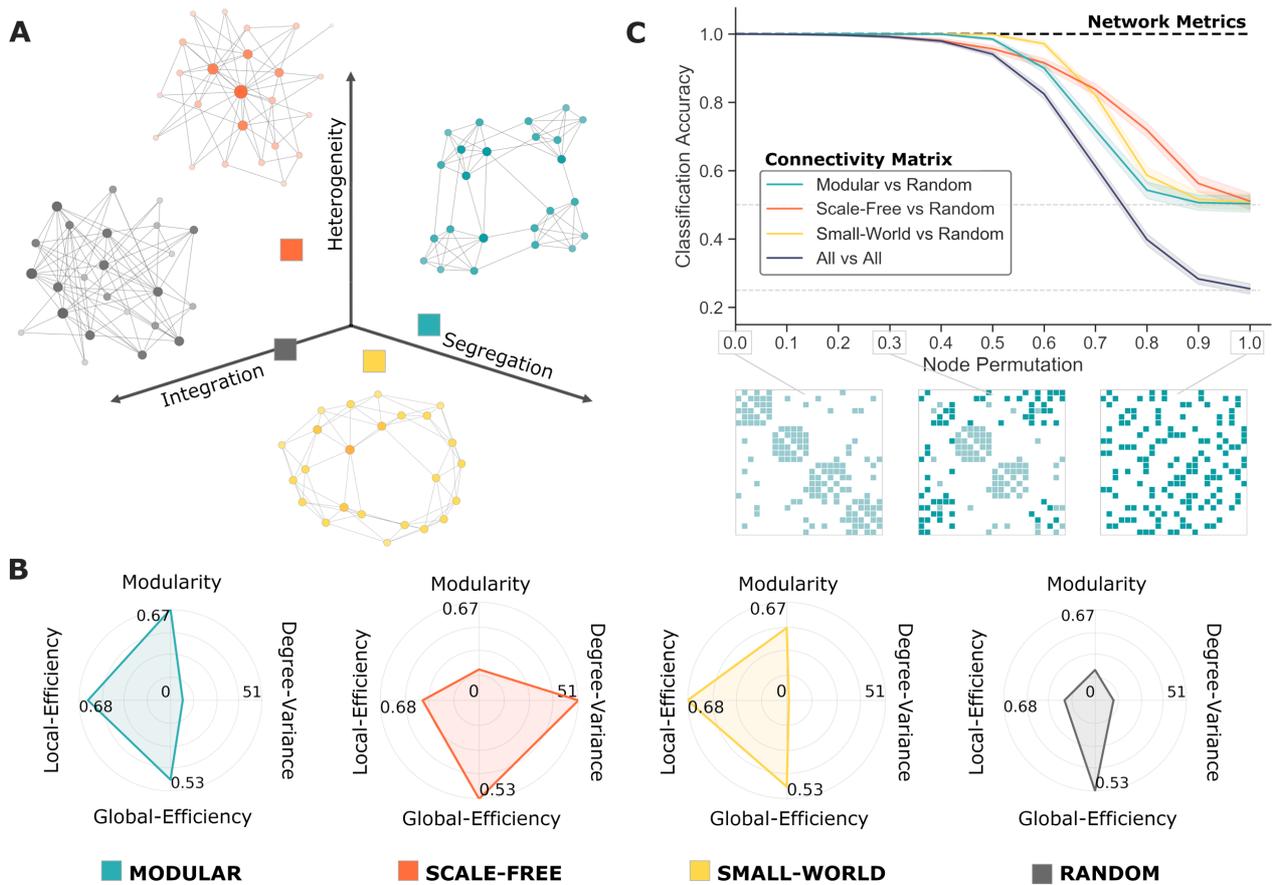

**Figure 2- Classification of networks via graph metrics.** Panel A) illustrates the graphs associated with each network class. For illustrative purposes graphs contain here N=24 nodes. Their position in the three-dimensional space qualitatively emphasizes their intrinsic properties in terms of segregation, integration and heterogeneity of information. Panel B) shows the radar plots for the mean values of the network metrics (Section 3.1) obtained from 1000 synthetic networks generated with different network models (**Box 2**). Each model corresponds to a different "class" of networks. Panel C) shows the accuracy results for the classification of the synthetic networks. Notably, 2-class and 4-class scenarios were performed, which is in line with the typical number of mental states used in BCI applications. Both connectivity matrices and network metrics were fed separately as input features into the classifier. Specifically, connectivity matrices were vectorized taking into account only the upper triangular matrix. Thus, the size of the feature vectors was 4950 for connectivity matrices and 4 for network metrics, respectively. To deal with the resulting complexity, we used singular value decomposition-based linear discriminant analysis (LDA) classifiers, which implement appropriate dimensionality reductions. To challenge the classifier, we increasingly permuted in a random fashion the nodes in the connectivity matrices. This corresponded to an increasing ratio of random relabeling of the nodes in the networks (x-axis). The line plots show the average value of the classification accuracy, while standard deviation is represented as shading patches around the average (obtained from a repeated random sub-sampling validation). For illustrative purposes, we also show an example for a modular network, where the darker colors correspond to the links of the nodes which have been permuted.

low computational cost and dimensionality, as well as in an easier direct interpretation.

The first study using network metrics in MI-based BCI classification was Daly et al. [234]. Authors assessed the discrimination ability of mean clustering coefficient to differentiate between tap and no-tap, in real and imagined finger tapping task. They recorded EEG data from twenty-two subjects performing the different task modalities. Then, to model the dynamics of inter-regional communication within the brain, they built functional connectivity networks by setting up phase synchronization links between each pair of electrodes. This resulted in a set of variable networks across time and frequency, potentially analyzable via graph theoretic tools. In order to characterize this synchronization dynamics, they computed mean clustering coefficients over the whole collection of networks. The result was a time-frequency map of mean clustering coefficients for each trial. The statistically significant differences between conditions, tap versus no-tap, suggested the potential of using time series of clustering coefficient as classification features. Thus, satisfying the fact that these features are not temporally independent, they used Hidden Markov Models (HMMs) to model and classify the temporal dynamics of these patterns. The discriminatory capability was superior when compared with traditional band power-based features, achieving accuracies above 70% for all subjects, which was not reproduced by band power approach.

Based on the same variations in phase synchronization during MI, Filho *et al* [235] also tested the potential of graph metrics to characterize these changes. In an offline study, EEG signals were recorded from eight participants during imagination of right and left hands movements using 64 electrodes. In the same direction as [234], networks were built for every 1 second window of left and right MI, but in this case they filtered the time series in two frequency bands of interest, $\alpha$ and $\beta$ bands. Then they computed five different graph theory metrics and used them as inputs for a least-squares based linear discriminant analysis classifier (LSLDA). At the same time, they extracted power spectrum density (PSD) features for a fair comparison. Using a leave-one-out cross-validation, the accuracies for single network metric classification were substantial, being around 80%, but when compared with PSD estimates its results were superior, being closer to perfect rate (99%). Nonetheless, the authors proposed a pair-wise combination of metrics which was enough to reproduce similar rates reached by PSD. Notably, the performance achieved by combined metrics involved a significantly smaller number of features, due to a selection of electrodes according to its individual classification rates. It is important to highlight that during the classification process this would be translated into less computational cost, which is encouraging when considering the implementation of network features in real time applications.

With a similar dataset, Uribe *et al* [236] investigated the potential of centrality measures to discriminate between left and right hand MI. They considered the difference between each pair of symmetric electrodes across hemispheres for every graph metric. They used degree, betweenness and eigenvector centrality to provide information regarding node's importance within the network. Two different classification methods were implemented, LDA and EML (Extreme Learning Machine), and feature selection was likewise based on classification rate improvement. Their results, expressed in terms of average classification error, showed better performance in $\alpha$ band when using degree centrality and EML. In a more ambitious attempt, the authors tested their approach on the BCI Competition IV 2a database. Using a wrapper feature selection their results were ranked in the third place, while the best known performance was obtained with PSD and CSP (Common Spatial Patterns) feature selection [237].

The introduction of network-based BCI should not necessarily imply the exclusion of traditional features. Instead,

it should be seen as a complementary approach to improve performance by integrating multiple neuronal mechanisms. In Cattai *et al* [238] they proposed different types of features combination. After revealing brain signal amplitude/phase synchronization mechanisms during EEG-based MI vs rest tasks, authors detected specific brain network changes associated with MI. Based on these findings, they computed spectral-coherence and imaginary-coherence connectivity matrices. The computation was performed for frequency bins in the 4 to 40Hz band with 1 Hz resolution, considering 9 electrodes in the sensorimotor area contralateral to the imagined movement. For every MI and rest trial, they extracted three types of features, coherence-based node strength, imaginary coherence-based node strength and power spectrum density. Then they tested all possible combinations with a cross-validated LDA classifier. While single node strength discriminations gave poorer results than power spectrum, their combination led to classification improvements in most of the subjects.

Zhang *et al* [239] also demonstrated the success of multimodal features fusion. Their cross-validated classification showed that the combination of node strength, or clustering coefficient, with CSP power selection, achieved higher accuracy than single feature. Getting accuracies over 70% for certain subjects. Noteworthy they chose the participants relying on their PSD-based MI-BCI inefficiency, i.e. its accuracy is under 70% [240] when using power spectrum. Similar to the previous study, it is interesting to point out that they also used spectral-coherence as a connectivity estimator. Their frequency selection was reduced to $\alpha$ band and, in order to avoid volume conduction effects, they selected 20 spatially separated electrodes. This is a potential explanation of the fact that they even got better accuracies than CSP when using single network metrics.

In a recent study, Gu *et al* [241] explored lower limbs MI. They did a detailed analysis of synchronization patterns in $\alpha$ and $\beta$ rhythms, to distinguish between left and right foot MI. Their study revealed a subset of sensorimotor networks exhibiting a cortical lateralization in the $\beta$ band with the respect to the imagined movement. Then the assessment with multiple network metrics showed a dynamic behavior between integration and segregation across each task repetition. Exploiting these results, they used and compared three variations of sparse logistic regression (SLR) to perform feature selection combined with support vector machine (SVM) classifier. The best accuracy was up to 75%, with all participants scores above the chance level, which is notable for foot MI discrimination. Furthermore, they contrasted the classification accuracy with features extracted with CSP method, but results were not able to outperform those obtained with network metrics.

As seen in section 4, network analysis can also be implemented in the study of other mental processes commonly evoked in usual BCI tasks, as for example cognition. In a preliminary study conducted by Buch et al. [242], a single subject with 122 intracranial EEG electrodes performed a test where reaction time was studied as an index of cognitive assessment. The experimental procedure consisted in a randomly chosen waiting period followed by a go signal after which the subject had to indicate its perception with a keypress; defining the reaction time as the delay between these two. Their premise was that dynamic changes in functional brain networks before and after the cue, could reflect temporal expectancy. Thus, they measured Phase-Locking Value (PLV) from sliding 500ms windows for the high gamma activity (70-100 Hz) of all pairs of electrodes, i.e. nodes, constituting the weighted links between them. They found that for fast reaction time trials, the immediate pre-cue period network (500ms before the cue) was characterized by a high node strength value compared to slower reaction times. When contrasting with traditional spectral based features, they did not find any pattern associated with reaction time variations. Going deeper in the network analysis, they computed communicability and showed a potential prediction ability based on the significant correlation between fast reaction time and high communicability in the left anterior cingulate. Motivated by these results, a SVM classifier was trained to discriminate between fast and slow trials, and then evaluated with a 10-fold cross-validation and permutation t-test. More precisely, they arbitrarily generated 2-classes labels and then randomized them 1000 times to create a null distribution of area under the curve AUCs. Results exhibited a reliable performance of the classifier (AUC = 0.72, p = 0.03). These results demonstrate the potential of network features as control signals for alternative cognitive-based BCIs.

Taken together, these results indicated the potential of network metrics as complementary features in BCIs. Future works should assess the robustness of these new features during online and real-time classification scenarios, where the reliability of the estimated brain networks becomes more challenging.

# 6. Conclusion

What does network science add to the dialogue between brains and machines? Far from being a fashionable tool, network science offers a grounded framework to analyze, model and quantify functional brain organization. On the one hand, network approaches can be used to understand how brain-computer interactions alter neural functioning on multiple temporal scales and which are the learning processes subserving BCI skill acquisition. On the other hand, one could use network science to extract innovative relevant features from functional brain networks to enrich the mental state characterization and improve BCI performance and usability. Network neuroscience offers therefore a theoretical backbone on which to begin testing hypotheses about the brain mechanisms of neurofeedback as well as developing and integrating such mechanisms into advanced BCI pipelines. In addition, the concepts that we have discussed throughout this review motivate and encourage future efforts that explicitly marry network approaches to machine learning in an effort to establish formal relationships.

In summary, we have provided an overviewed tutorial on what a network actually is, how to characterize it and how to use it in BCI and neurofeedback-related scenarios. While we have focused this review on the application to noninvasive electrophysiology data from adult humans, particularly collected during motor imagery-based tasks, we anticipate that the same approach will also be of interest in future applications including different neuroimaging techniques (invasive or noninvasive), BCI paradigms (synchronous or asynchronous), as well as brain diseases.

One crucial aspect to take into consideration when conceiving online BCIs is the time complexity associated with the computation of network metrics. Despite all the implemented algorithms are bounded by a polynomial time complexity, this should be carefully considered when measuring the overall expected latency of BCIs [243]. Another important issue is the statistical uncertainty associated with the estimation of functional edges of the brain network. Indeed, researchers are often obliged to arbitrarily filter a number of connections in the network, while in principle network metrics can be applied to fully connected and weighted graphs. We believe that more research is needed in this direction to obtain more reliable connectivity patterns reflecting true interactions in the brain.

In conclusion, our aim is to provide the neural engineering community with both tools and intuition, and to support the growing interest in ameliorating BCIs through network neuroscience.

## Acknowledgements

Authors would like to acknowledge Vincenzo Nicosia for useful suggestions and comments on the computational complexity of the introduced network metrics. The content is solely the responsibility of the authors and does not necessarily represent the official views of any of the funding agencies.